# Electron and lattice structure of ultra thin Ag films on Si(111) and Si(001)

V.A. Gasparov[a] and M. Riehl-Chudoba[b]
[a] *Institute of Solid State Physics RAS, 142432, Chernogolovka, Moscow district, Russian Federation*
[b] *Institut für Festkörperphysik, Friedrich-Schiller-Universität Jena, Max-Wien-Platz 1, 07743, Jena, Germany*

We studied the low temperature (T≤130K) growth of Ag on Si(001) and Si(111) flat surfaces prepared by Si homo epitaxy with the aim to achieve thin metallic films. The band structure and morphology of the Ag overlayers have been investigated by means of XPS, UPS, LEED, STM and STS. Surprisingly a ($\sqrt{3}\times\sqrt{3}$)R30$^o$ LEED structure for Ag films has been observed after deposition of 2-6 ML Ag onto a Si(111)($\sqrt{3}\times\sqrt{3}$)R30$^o$Ag surface at low temperatures. XPS investigations showed that these films are solid, and UPS measurements indicate that they are metallic. However, after closer STM studies we found that these films consists of sharp Ag islands and ($\sqrt{3}\times\sqrt{3}$)R30$^o$Ag flat terraces in between. On Si(001) the low-temperature deposition yields an epitaxial growth of Ag on clean Si(001)2x1 with a twinned Ag(111) structure at coverage's as low as 10 ML. Furthermore the conductivity of few monolayer Ag films on Si(100) surfaces has been studied as a function of temperature (40-300 K).

## 1. Introduction

Most semiconductor surfaces reconstruct after metal deposition onto clean surfaces which is subject of many investigations. In particular, the Ag/Si(111) and Ag/Si(001) surfaces have been studied by almost every surface analysis techniques and many different surface phases (SP) have been observed for monolayer and sub-monolayer coverage's (see [1-24] and references therein). At larger coverage's however, the metals are known to form 3D-cluster structures at RT due to the large lattice mismatch with respect to Si substrate. In general the metal lattice parameter is smaller than those of the semiconductors. The hetero epitaxial growth of the metal films on semiconductors is a topic of great fundamental and technological interest [1,2], in particular due to its potential in the development of quantum devices.

Recently, a breakthrough has been achieved in this field by use of low substrate temperatures during the growth of metallic films [2-12]. The high mobility of metal atoms at RT is tremendously diminished at low temperatures which overcomes the tendency toward islanding. In fact, the layer-by-layer growth of Au as well as Ag films on Si(111) at 95K was concluded as evidenced by RHEED intensities oscillations as a function of coverage [10-12]. This 2D growth mode of Ag on Si(111) at low temperatures has been confirmed by low-temperature STM investigations [3]. However, small 2D Ag islands (20-40Å size at 4-5 ML coverage) rather than continuous films have been observed. Interestingly a layer-by-layer growth of Pb films at low temperatures was evidenced by RHEED oscillations when the Si(111)7x7 substrate was modified by sub-monolayer Au deposits (Si(111)$\sqrt{3}\times\sqrt{3}$Au or (6x6)Au structures) prior to the Pb deposition [11,12]. Similar results have been obtained for Au and Ag growth on Si(111)$\sqrt{3}\times\sqrt{3}$Ag (hereafter $\sqrt{3}$Ag/Si) and Si(111)$\sqrt{3}\times\sqrt{3}$Au substrates, respectively [13,14]. It was concluded that the epitaxial growth of these noble metals is promoted by the formation of a 2D Ag wetting layer on the Si(111) surface. Furthermore, the continuous growth of Ag(111) films on Si(001) at low temperatures has been observed by SPA-LEED oscillations and STM investigations [4-9]. It is reported that the electrons are confined in metallic islands due to the lack of translational symmetry thus producing quantum well states (QWS) [15-18]. These states begin to appear as the thickness of a 2D layer becomes comparable to the de Broglie wavelength of electrons and have been observed by photoemission studies [15-18].

Despite the wealth of obtained information, the details of the electron structure of √3Ag/Si SP are still controversial. In fact, recently it was shown that this superstructure exhibits an isotropic and parabolic surface state band crossing the Fermi level, which gives rise to a 2D metallic like system [2,19]. The electrical conductivity of a Si(111)7x7 and √3Ag/Si surface was measured in UHV by a microscopic four-point probe method [20]. Clear evidence of an unusual two order of magnitude enhancement of surface conductivity by deposition of 1ML √3Ag/Si SP and reducing the probe space from *mm* scale to the *micron* one was obtained. Even though some experimental evidence of electrical conduction through √3Si/AgAg/Si SP band by aid of four-point microprobe *mm* and *micrometer* scale experiments at room temperature has been described [2], this result is in contrast to the observation of nonmetallic behavior of √3Ag/Si SP at low temperature measurements [4,5]. The disadvantage of these conductivity experiments performed at room temperatures is that the surface conductance is screened by the substrate. Therefore low temperature experiments are crucial for an ambiguous conclusion.

In our group several studies of conductance and structural properties of metal films (Cr, In, Pb) on semiconductor surfaces have been performed. We reported on the growth and the *ex-situ* temperature dependence of the electrical conductivity and the magnetoresistance tensor components of the Si(111)-Cr(√3×√3)R30°, Si-In(1×1) R30° surface phase and of ultrathin Si-Cr films covered by a thin amorphous αSi protecting layer [21,22]. We found that Si-In-αSi SP is a 2D semiconductor with an enhanced mobility, while Si-Cr-αSi SP behaves metallic with a rather low mobility. Lately, the *in-situ* temperature dependence of conductivity of the Si(111)-7×7 and SP of In and Pb on clean Si(111) surfaces prepared at room temperature have been studied [23,24]. It was shown that both Si-Pb-αS and Si-In-αSi SP exhibit semiconductor 2D behavior with a very small mini gap (0.09meV and 0.0035meV, respectively). The percolation behavior and metal-insulator transition was observed for In [23] and Pb [24] island films on Si(111).

Recent experiments [2,4,5] show that 2D conductivity may be realized in Ag superstructures on Si. The measurements of the resistivity of epitaxial metal films are very important for the study of an ideal 2D metallic system. Unfortunately, these studies of conductivity were performed on samples with no *in-situ* control of the surface structure, such as STM. Amorphous and granular films may show rather peculiar effects like a change of the sign of temperature coefficients of the resistance. Therefore we performed new experiments for LT deposition of Ag films and LT conductivity measurements combined with electron structure and morphology investigations of these films.

In this paper we describe XPS, UPS, LEED, STM and STS as well as the electron transport measurements that show very peculiar properties of Ag superstructures on Si(111) 7x7 and Si(100)2x1 atomically flat surfaces prepared by homo epitaxy. The low temperature growth of Ag thin films on preliminary prepared √3Ag/Si SP results in epitaxial metallic films on Si(111) substrates according to XPS, UPS and LEED studies. However, STM investigations indicate sharp Ag island formation with √3Ag/Si layers in between. For the Si(100)/Ag case, we observe 2D-LT growth of continuous metallic flat Ag(111) films. A remarkably high conductivity for these films is found which decreases as a function of temperature between 100 and 45 K, where the substrate becomes almost insulating. By using well prepared substrates at low temperatures we were able to measure almost pure surface state conductance even over macroscopic distances.

## II. Experimental

The growth experiments have been performed in an UHV multichamber surface analytical system with a base pressure of $2\times10^{-8}$ Pa, which allows *in-situ* preparation of the Si surfaces as well as characterization by means of UPS, XPS, AES, LEED and STM [25,26]. The UHV chamber was equipped with a water-cooled Knudsen boron nitride effusion cell for Si homo epitaxy. The deposition rate was 1ML/min (1ML =$7.8\times10^{14}$cm$^{-2}$, the site density of the unreconstructed Si(111) plane). The STM is a commercial large sample type from Omicron. Electrochemically etched (2M NaOH) single crystalline W<100> and hand cut PtIr tips have been used which were out gassed and flash annealed in UHV. The images were recorded at room temperature in the constant current mode at voltages ranging between -3.0 to +3.0 V and a current between 0.01 nA - 0.05 nA. All given voltages are referring to the substrate polarity, the tip was grounded. Scanning Tunneling Spectroscopy (STS) spectra were obtained by measuring the tunnel current *I* versus the voltage *V* across the tunnel junction.

The resistivity experiments were performed at a base pressure of $1\times10^{-8}$ Pa in a LAS-630 spectrometer equipped with AES, LEED and a home made variable temperature (40-1200 K) sample holder. A four contact geometry was used with pre-deposited Mo contact pads on identically prepared Si(100)2x1 samples (5x5 mm$^2$), that were separated by slits machined into the samples by a diamond saw [23,24]. Thus the effective examined sample size was 2x2 mm$^2$. The surface conductivity was measured according to the van der Pauw technique as reported before [23,24].

The substrates were n-type P doped Si(001) and Si(111) wafers of 5 $\Omega$ cm resistivity, at RT oriented within 0.1º to the <111> and <001> directions, respectively. The resistivity study has been performed on a n-type Si(001) substrates of 1 k$\Omega$ cm at RT. Atomically clean surfaces were prepared as follows: degreasing in acetone and methanol, etching in NH$_4$F/HF and rinsing in deionized water. Thus, a hydrogen terminated surface was obtained. The samples were loaded into the UHV chamber through a load-lock and out gassed by step annealing from 300°C up to 600°C for 12 h. Then the samples were flashed to 1200°C several times in order to remove native oxides and cooled down rapidly to 600°C, keeping this temperature for 10 min before it was allowed to cool to RT within about 30 min. The samples were fixed on a molybdenum holder and could be heated by means of electron bombardment of the rear side of the sample up to 1200°C and cooled down to 130 K by a liquid nitrogen flow cryostat. The temperature was monitored with an AuFe - Chromel thermocouple pressed against the front face of the wafers. The pressure was maintained in the 10$^{-7}$ Pa range during annealing and flashing. After this cleaning procedure the surface exhibited a bright and well defined LEED patterns according to the Si(001)2x1 and Si(111)7x7 reconstructed surfaces, respectively.

The STM experiments were performed at a base pressure of $5\times10^{-8}$ Pa. They were showing a rather rough surface with small (7x7) domains, even though a good 7x7 LEED structure was observed. Therefore, unlike to other studies, an epitaxial Si film of about 10 nm thickness has been grown subsequently by the following two step process: (i) 5 nm Si evaporation at a temperature of 450°C and (ii) 5 nm at a temperature of 650°C with a growth rate of 0.1 ML/min in order to obtain a large scale atomically flat surface. LEED and the STM consistently verified well-ordered Si(111)7x7 and Si(001)2x1 structures. These surfaces were regarded as atomically clean owing to the complete disappearance of C and O peaks in the XPS spectra. Subsequently, an epitaxial Si film of about 10 nm thickness has been grown at a temperature of 650°C, which resulted in an atomically flat Si(111) (7x7) surface with a typical terrace width of about 400 nm.

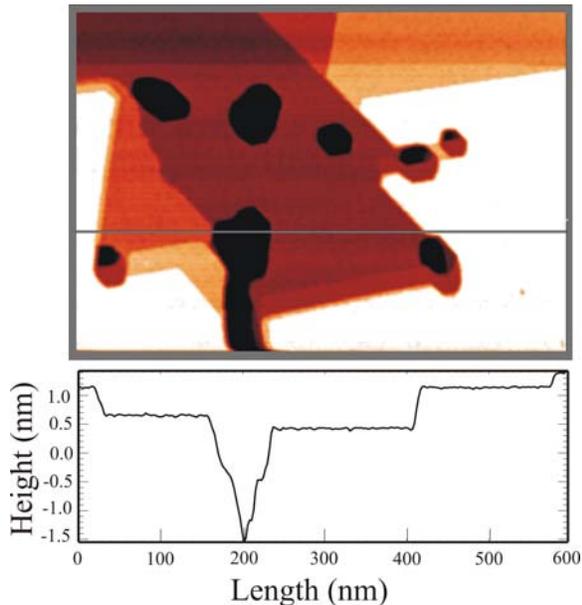

FIG.1. (Color online) Typical 600×600 nm STM image of Si(111) (7×7) reconstruction prepared by homo epitaxy. Black holes likely correspond to surface defects.

Ag overlayers have been achieved by deposition from a tantalum crucible at a rate of 0.01 nm/s onto the Si substrates held at 500°C. The mass flux was monitored by a quartz–crystal microbalance and the Ag coverage was determined by the exposure time of the evaporation. The deposition rate was 0.1 ML/min, where 1ML corresponds to the surface density of Ag(111) equal to $1.38 \times 10^{15}$ atoms/cm$^2$.

## III Results
### IIIa. Si(111)√3x√3Ag Surface

A typical topographic STM image of a homo epitaxially prepared Si(111)7x7 surface is shown in Fig.1. The STM topography reveals large (200 nm) atomically flat regions of well ordered Si(111)7x7 phases with atomic steps in between. In order to characterize the structures of √3Ag/Si, we performed LEED investigations for these films. Figure 2 shows the diffraction pattern around the Si(10) diffraction spot for a √3Ag/Si surface, a 5 ML Ag film deposited at 130 K on this surface as well as for a 10 ML Ag film evaporated at LT on Si(100)-(2x1) (Fig.2c).

A large area topographic STM scan of a √3Ag/Si surface prepared at 500°C is shown in Fig. 3. In this image, reflecting the filled states of the sample, the position of √3Ag/Si islands appear as bright areas. A high-resolution filled-state STM scan recorded at RT on these islands is shown in Fig.3b. Atomically resolved STM images indicate the characteristic honeycomb structure with a low defect density. LEED as shown in Fig.2 confirm a sharp (√3x√3) structure and XPS showed no detectable surface contamination. The atomic arrangement of this well known √3Ag/Si superstructure has been solved by the honeycomb chained trimer (HCT-1) model [27].

The detailed STM study showed wide atomically flat terraces with 2D islands of double mono atomic step height, indicating the partial second layer of √3Ag/Si on the Si substrate [26]. These non percolated terraces are connected by a non conductive wetting layer which is responsible for the percolation resitivity on this surface [4,5]. The profile lines of the STM images and STM nano structuring experiments on the same surface [26] have shown that the average height of these islands is 3.1 Å above the adatoms of the wetting layer. This is significantly larger than the Ag(111) step height of 2.36 Å. Thus formation of √3Ag/Si islands (Si trimer with Ag layer on top) at coverage of 1 ML silver showing a uniform thickness of one atomic layer was observed. The lateral size of these islands varied from 60 nm to 600 nm, indicating that once a √3Ag/Si island is formed, it starts to grow in size laterally. We did not find the formation of 3D Ag clusters on √3Ag/Si SP which has been observed at higher Ag deposition temperatures [29,30]. Notice, that Su *et al.* [28] and Sobotík *et al.* [29,30] observed that flat Ag islands with (111) orientation of 2-layer thickness can actually grow on the Si(111)7x7 surface even at room temperature. The 2D growth behavior of 2-layer islands is likely driven by the QWS [28-30].

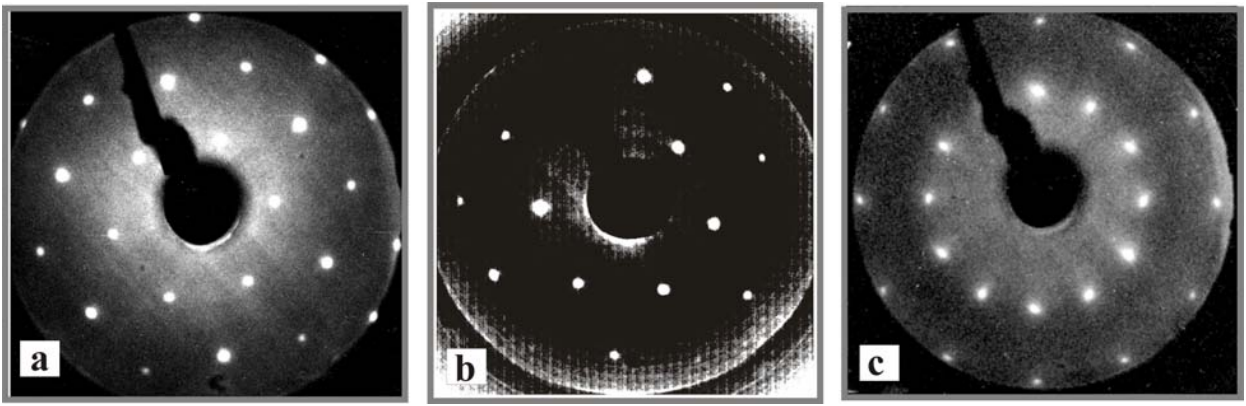

FIG.2. Typical LEED patterns of: a) √3Ag/Si surface, b) 5 ML Ag deposition at 130 K on this surface, c) 10 ML Ag film evaporated at LT on Si(100)-c(2x1) surface, recorded at RT. The electron beam energy are 48 eV, 85 eV and 130 eV, respectively.

In order to study the local electron structure of the √3Ag/Si surface we performed STS experiments simultaneously with STM imaging under the condition of a fixed tip-sample distance, according to Ref. 25. That is, a measurement of a tunneling current versus bias voltage spectrum with the STM feedback loop turned off at each point during scanning of the tip over the surface. From these spectra we can calculate the normalized differential conductivity *(dI/dV)/(I/V)* which is proportional to the local surface density of states (LDOS) on the topographic image. The Fig. 4 shows normalized differential conductivity, as calculated numerically from the *I-V* curves, averaging the profiles of the STM image of √3Ag/Si islands. In this figure the Fermi level corresponds to 0 V and the energy is defined by the sample voltage relative to $E_F$.

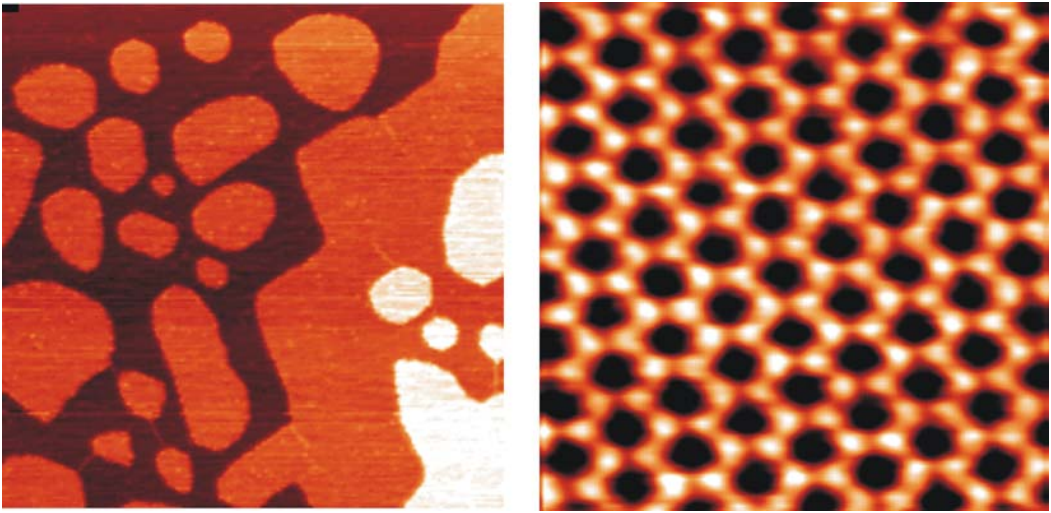

FIG.3. (Color online) STM topographic image of √3Ag/Si at 265×265 nm$^2$ (left panel) and at high resolution 12 x 12 nm$^2$. The images were recorded at 0.1 nA tunnel current and -0.6 V bias voltage. resolution 12 x 12 nm$^2$.

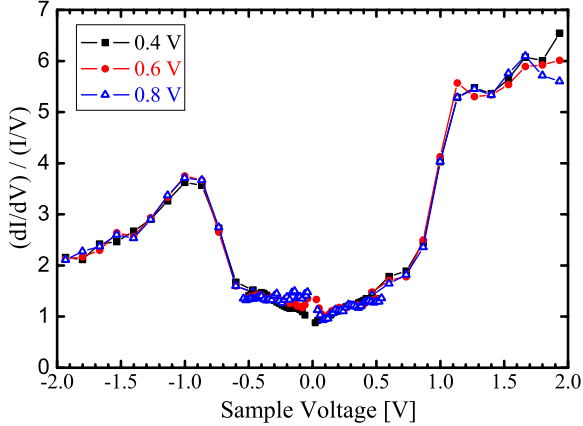

FIG.4. (Color online) The normalized differential conductance (dI/dV)/(I/V) vs V STS spectra from the *I* vs *V* data measured for different tip-surface separations.

When comparing the normalized differential conductance *(dI/dV)/(I/V)* curves of surface sites, we see similarities for different islands at different tip-surface distances determined by bias voltage. Of course a large difference in the differential conductance, *dI/dV*, for different bias voltages was observed. A significant feature of these dependencies is the observation of two maxima at -1 eV and +1 eV in the normalized differential conductance as presented in Fig. 4. The peak positions remain unchanged for different islands indicating a uniform LDOS distribution in the area where the √3Ag/Si structure is formed. Apparently, it appears a reduction of LDOS below and above $E_F$ compared to the 7x7 surface. The *(dI/dV)/(I/V)* in the same range of V is much smaller on the √3Ag/Si islands than on the Si(111) 7x7 surface. Thus a semiconductor band structure of the √3Ag/Si surface with an energy gap of 2 eV is observed in contrast to 2D metallic like system as proposed in Ref. [2,19]. Notice usual STS problem because the both measured conductance and differential one drop below the noise level close to 0 V. This is due to plateau region of the *I-V* curves inside a gap. Therefore, calculated value of *(dI/dV)/(I/V)* is very noisy close to 0 V due to ratio 0/0 [25]. In order to avoid this problem, we fitted *I* vs *V* dependence close to zero voltage and calculated *(dI/dV)/(I/V)* from this fitted curves, that is what is shown in Fig.4. Note also, that metallic behavior will produce linear *I* vs *V* dependence, while we have a plateau.

We performed UPS on √3Ag/Si in order to check the consistency with the STS results regarding the band structure of √3Ag/Si. The Fermi – level position was determined by photoemission from the Mo sample holder. Figure 5 shows the angular-integrated valence-band photoemission spectra of the √3Ag/Si surface and a 5 ML Ag film evaporated at 130 K on this surface. The spectra are taken at 70° incidence of the photon beam with respect to surface. As a reference we measured and included a scaled UP spectrum of an Ag bulk single crystal. The spectra are shifted along the vertical axis for clarity. Spectral features of all these surfaces are quite different, while the data for the 5 ML Ag film and the bulk Ag crystal are rather similar.

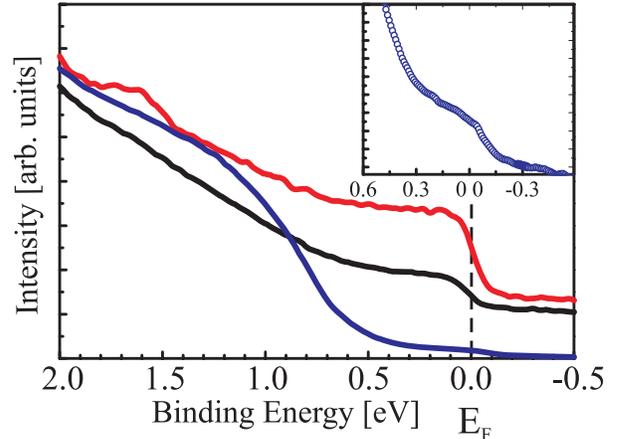

FIG.5. (Color online) The angle-integrated valence band UP spectra for √3Ag/Si surfaces and for a bulk Ag single crystal (upper curve). The lower spectrum was measured on a regular √3Ag/Si surface while the middle curve was obtained from 5 ML Ag film on √3Ag/Si surface. The photon source was He I (hν=21.2 eV). Inset shows extended part PE spectra close to $E_F$ for √3Ag/Si surface.

When comparing the normalized conductance STS curves with UPS data, we see similarities for the filled states. As in STS experi-ments a sharp reduction of the DOS at binding energy $E - E_F = -1$ eV can be seen in the PE spectra of √3Ag/Si surface in Fig.5. These UPS data at -1 eV are in a good agreement with $S_2$ states observed in ARUPS experiments [15]. We did not find metallic $S_1$ states in our STS studies likely due to the highly local character of the STS experiments performed only on top of √3Ag/Si islands, in contrast to sample averaged ARUPS or UPS data. This suggests that the surface band structure of √3Ag/Si surface is inherently semiconducting and metallic behavior observed from ARUPS [2,19] may be due to Ag excess on top. This observation can also explain the observed band banding close to the surface [19]. From the comparison to the bulk Ag PE spectrum and a 5 ML Ag film evaporated at 130 K on √3Ag/Si, we can see a quite clear Fermi edge for the LT prepared films being consistent with the metallic character. Subsequent annealing of the LT prepared structure up to 500°C resulted in the same spectra. At the same time, the √3Ag/Si surface exhibit clear semiconducting behavior with the same gap for the filled states as in STS experiments. Apparently, the tiny Fermi edge formation is due to Ag excess on top of √3Ag/Si islands as in Refs.31,32 (see inset to the Fig.5).

It is currently believed that the √3Ag/Si surface is arranged according to the HCT-1 model [27]. In this model the top Ag and uppermost Si atoms form trimers on the surface. However, recently, different versions of this model have been suggested for the LT phase of √3Ag/Si surface with a slightly asymmetric modification of the Ag trimers described by the in equivalent triangle (IET) model (see references in [2]). This model was proposed for the explanation of the asymmetric appearance of √3x√3 unit cells observed at 62 K as compared to those of the √3Ag/Si surface observed at RT [2]. Accordingly cooling from RT to LT should provoke a phase transition from HCT-1 to the IET surface. Theoretical calculations based on the HCT-1 model of the surface band structure reveal that the bands are degenerated at the $K$ point of the surface Brillouin zone, while the bands are split within the IET model [28]. However, recent LT ARPES studies of Uhrberg *et al.* of the √3Ag/Si surface did not find any splitting at low temperatures [31]. Therefore, it was concluded that the HCT-1 model for the √3Ag/Si surface is the most reasonable one.

Also it was proposed from ARPES studies that the √3Ag/Si surface is intrinsically metallic [2,19]. Some indication of a partial surface metallization can be seen from our UPS data in the inset of Fig.5 indeed. However, recent ARPES studies [31,32] have shown that there is a significant effect of additional Ag atoms on the surface electronic structure near the Fermi level. A few percent of a ML of additional Ag atoms adsorbed on top of √3Ag/Si, results into a occurrence of partially occupied surface states. These states are completely removed for clean √3Ag/Si without any additional Ag on top. Although the difference between UPS and STS data is clear, we have to point out that islands structure of √3Ag/Si surface may manifest the photoemission data, while the STS data in Fig.4 are taken from well defined regular √3Ag/Si surface sites. Thus we believe that the tiny indication of the Fermi edge observed in Fig.5 is due to additional Ag atoms adsorbed on top of √3Ag/Si and the pure √3Ag/Si surface is semiconducting according to our STS data (see Fig.4).

Another important result of our UP data is the observation of a clear metallic behavior of the 5 ML Ag film on top of the √3Ag/Si surface. In order to decide whether these films are solid, we performed XPS studies with the same light incident angle of 20° with respect to the surface normal as in the UPS experiments. We found that the general shape of the XPS spectra for the pure √3Ag/Si superstructure is quite similar to the published ones. These spectra show peaks that can be related to the original components of Ag and Si of the surface atoms with no indication of contaminations. After the deposition of 5 ML Ag at 130 K the XPS data

Ag at 130 K the XPS data show that the Si 2*p* peaks almost disappear,

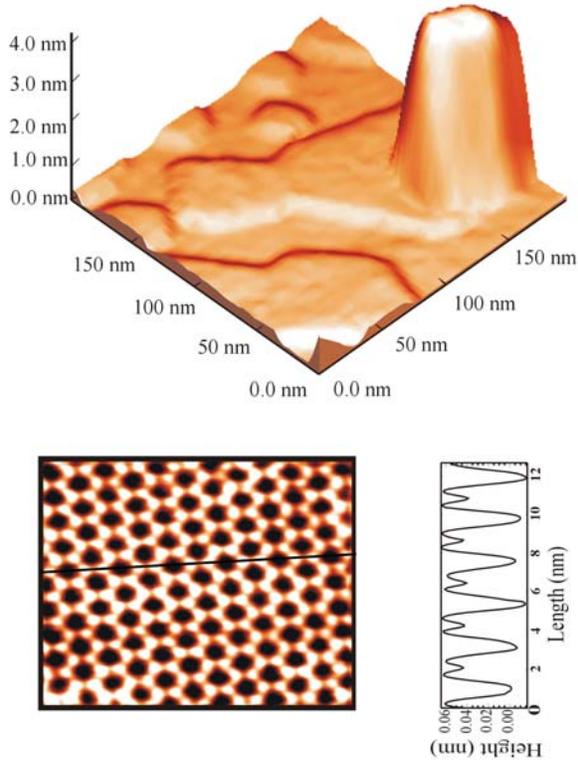

FIG.6. (Color online) Filled state STM image from 5ML Ag deposited at LT onto √3Ag/Si showing large Ag islands with √3Ag/Si areas in between (top panel). The atomically resolved STM image recorded in between the Ag islands (9×12 nm$^2$) indicates the √3×√3Ag structure (bottom panel).

indicating that theAg films on the √3Ag/Si substrate are almost solid. Thus, according to our LEED, UPS and XPS studies, we may conclude that the low temperature evaporation of 5 ML Ag films on √3Ag/Si leads to an epitaxial growth of metallic solid films.

In order to confirm or not this unusual conclusion we performed STM studies of the 5 ML Ag films evaporated at LT on √3Ag/Si. We first prepared the √3Ag/Si phase by high temperature Ag deposition. Then we subjected the √3Ag/Si to LT (130 K) deposition of a 2-5 ML of Ag on top of this phase in order to achieve 2D films. Figure 6 shows an STM image of the √3Ag/Si surface after deposition of 5 ML of Ag films at 130 K and subsequently warmed up to RT. Contrary to the LEED and XPS data, we found that this film consists of rather narrow and large Ag islands with atomically flat Ag(111) top facets in between. The typical width of these islands is 50 nm and the height is about 4 nm.

The area between these islands retains atomically flat √3Ag/Si terraces with a width of about 200 nm. The symmetry and orientation of the flat terraces between these islands with clear honeycomb geometry suggest a √3Ag/Si structure in agreement with the HCT-1 model [27]. This conclusion is confirmed by the high resolution STM image with the profile presented in the lower panel of the Fig.6. This image reveals hexagonal features consisting of six protrusions related to Ag trimers on top while the dark spots apparently are related to the Si trimers.

The observed Ag island distribution may be explained by diffusion of the Ag atoms from the wetting layer across the surface while warming up to RT. Apparently, the surface free energy drives the system into the energetically favorable configuration characterized by larger isolated Ag 3D crystallites. The Ag atoms migrate to the top of existing islands during warm up to RT thus increasing their height and preserving their regular shape. A similar behavior has been observed for the transfer of Pb from the wetting layer of √3Pb/Si surface to the seven layer islands during annealing from 115 K to 180 K by low temperature STM experiments [33]. Therefore, additional LT STM studies of LT prepared Ag films on √3Ag/Si substrates with different coverage are needed in order to elucidate the exact atomic structure of these films and to see whether the islands size distribution is determined by QWS [28,34].

**IIIb. Ag(111) films on Si(001)**

First we prepared the Si(100)-(2x1) superstructure by high temperature homo epitaxy of Si as described for the Si(111) surface. The structure and composition of this surface was controlled by LEED, XPS and STM. We then subjected the Si(100)-(2x1) surface to the deposition of 5-20 ML Ag films at 130 K. Figure 2c illustrates the LEED pattern from the 10 ML Ag film after annealing to RT. From first glance the LEED pattern are very peculiar. The zero spot is surrounded by 12 first-order spots and 12 second-order ones. However, a closer analysis can easily explain these pattern as a superposition of the twinned domains of Ag(111) surface, i.e. a Ag(111) LEED pattern rotated by 90° with respect to each other [8,9]. The spots are elongated in two perpendicular <110> directions indicating superposition of two Ag domains rotated by 90°.

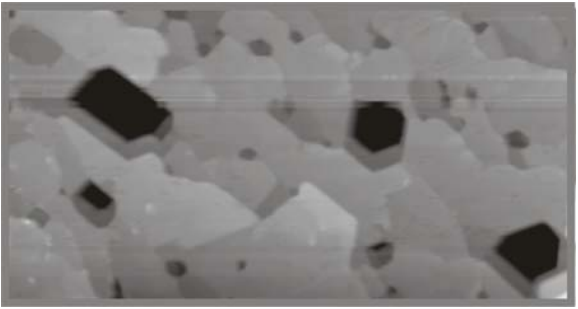

FIG.7. STM 300x160 nm² topographic image of 10 ML Ag film deposited on a Si(100)-c(2x1) surface at 130 K and annealed to RT. The topography was acquired at 1 V sample bias and 1 nA tunneling current.

The morphology of this 10 ML film was studied by STM and is shown in Figure 7. The cornered shape of the step edge confirms the in-plane orientation of the Ag grains as deduced from the LEED pattern. The reason for this observation can be understood by comparing the Ag and Si lattices. The Si/Ag interface exhibits an almost exactly 4/3 mismatch ($a_{Si}$=0.5431nm and $a_{Ag}$ = 0.4086 nm), thus the distance between three Si atoms matches the distance between 4 Ag atoms [8,9]. According to a recent reports ultra thin Ag(111) films are epitaxially formed by deposition up to 40 ML of Ag at LT (< 130 K) on Si(001) surfaces followed by post annealing to RT [8,9].

In order to understand whether these films are metallic we studied the electronic states on this surface by angle integrated UPS. Figure 8 shows the PE spectra for the Si(001)-(2x1) surface and the 10 ML Ag film grown on it at 130 K and annealed to RT. The spectra are taken at He I photon energy (hν=21.2 eV) and at 70° incident of the photon beam. The drastic change of the electronic structure of the Si(001) surface induced by the Ag film is obvious. The spectral shape of the valence band PE spectra from the 10 ML film is similar to those of the Ag deposited on top of √3Ag/Si and the Ag single crystal presented in Fig.5 indicating metallic state formation. The formation of such states for the LT deposited films is consistent with our STM data in Fig.7 and the LEED results shown in Fig.2c.

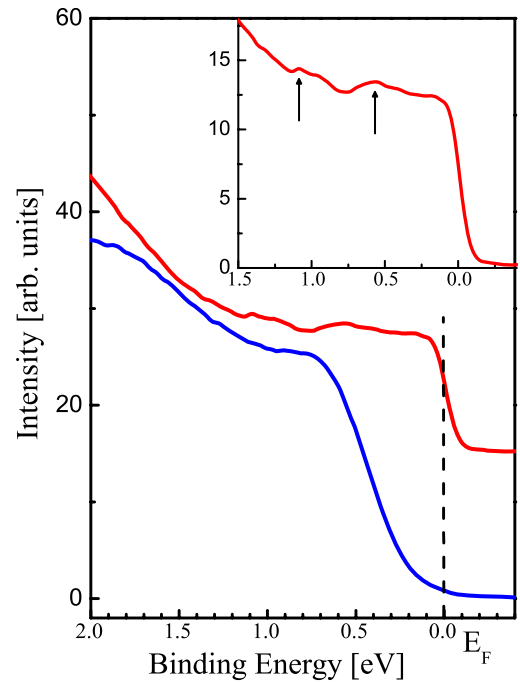

FIG. 8. (Color online) The angle-integrated valence band PE spectra obtained at RT from a Si(100)-(2x1) surface (bottom curve) and from 10 ML Ag film (top curve) deposited on it at 130 K. Inset shows extended part of PE spectra close to $E_F$ for this film.

Several small peaks at binding energies of 1.2 and 0.6 eV below the Fermi edge appear which may be assigned to QWS [18,34]. These states are due to energy quantization phenomena of confining electrons in a spatially narrow potential well produced by the thin film. In this case the thickness of atomically flat film has to be an integer of the Fermi wavelength. The thickness of the 10 ML Ag film is about 2.36 nm which seems to be enough to realize the quantum confinement of carriers in the film. Thus our observation of these QW states is another proof about 2D metallic character of the Ag films on the Si(100)-2x1 surface.

Finally we can conclude that the Ag film deposited at LT on Si(001)-c(2x1) becomes metallic with a 2D twinned Ag film after annealing to RT due to the integer-ratio lattice mismatch of 3/4 in agreement with Refs.7-9,18,34. At the same time Ag grows at RT on Si(100)-2x1 clean surface in island-growth mode up to 40 ML coverage [4].

### IIIc. Resistivity of Ag(111) films on Si(001)

Our first typical result of conductance versus temperature measurements of a Si(100)-2x1 substrate and a 4 ML Ag film on it deposited at 45 K are shown in Fig. 9. The conductance of the clean substrate sharply drops in the temperature range from about 170 K to 45 K monotonously due to freezing out of the number of carriers. In turn, the resistance drops in one order of magnitude after 4 ML Ag film deposition. Thus we can account for the conductivity of the Ag film by subtracting the substrate conductance. The result is shown by triangles and displays a completely different behavior as compared to the substrate. The film conductance has a maximum at 140 K and then drops at higher temperatures. Note however, that the data precision is rather low beyond this maximum due of the small conductance change with respect to the substrate. In contrast to the expected metallic behavior the conductance sharply decreases below the maximum but is not reaching zero as the substrate.

The observed difference in $\sigma(T)$ is obviously due to the different conduction mechanisms in the film and the substrate. For the weak localization and low temperature region a logarithmic temperature dependence of the conductance $\sigma(T) \propto \log(T)$ is expected. On the other side, according to variable range hopping conductivity within a band of localized states, $\sigma(T) \propto \exp(T_o/T)^{1/3}$ could be expected [35]. However, we could not fit the data quantitatively with any one of these laws. A reasonably fit was obtained for a modified variable range hopping conductivity with a non-zero residual conductance:

$$\sigma(T) = \sigma(0) + \sigma_1 \cdot \exp(-T_0/T)^{1/3} \qquad (1)$$

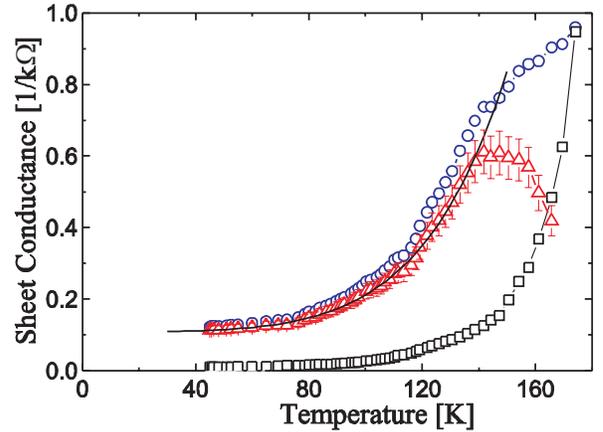

FIG.9. (Color online) The conductance versus temperature dependencies for a 4 ML Ag film (circles) deposited at 45 K on Si(100) and for the bare Si(100) substrate (squares). The triangles show σ vs T for the Ag film after substrate subtraction.

The fit with Eq.1 is shown in Fig.9 as a solid line through the Ag film conductance data. Thus we can speculate that the residual conductance might be due to in grain conductance while the second term is due to the conductance inside a grain blocked by the energy of charging up a grain. The grain boundaries are large angle boundaries within 90 degrees of in-plane rotation according to the structure model [8,9].

So far it is not clear whether this or other theoretical model can provide a satisfactory agreement with the observed temperature dependence of conductivity. We believe that additional studies for thicker films are needed ditional studies for thicker films are needed in order to solve the problem of ultra thin film conductance.

The relation between the surface structure of noble metal layers on top of Si(111) and their electrical conductivity has been emphasized in a number of papers (see [2]). Herein a new type of conduction through 2D electron systems localized in the surface-state bands was proposed. Although the results are really spectacular, there are conductivity problems. Simple estimations of surface conductivity as the difference from sample conductivity with superstructure minus substrate conductivity, shows that the changes observed can be due to too highly conductive surface layer with a conductivity $\rho_{Ag}$ of metallic silver if we consider one mono-layer thickness.

Apparently, the sharp metastable resistivity drop produced by metal evaporation at room temperature is caused by the adsorption-induced changes in surface states and band bending beneath the surface rather than the conductance through the surface-state bands. Notice also, that recently the rapid drop of resistance was observed in similar experiments only after the critical coverage of 1-3 ML [37], while earlier [36] this drop took place at sub monolayer coverage. This phenomenon was associated with an island coalescence and formation of percolation paths for electrical conduction [37]. Indeed, our data as shown in Fig.3 and other STM studies for Ag on Si(111) surface [28-30], indicate that two layer thick Ag (111) or √3Ag/Si islands occur with large separations among them rather than the formation of a continuous 2D electron system localized in surface bands.

Our goal is the preparation of a real 2D metallic conducting layer on silicon with enhanced mobility and we believe that Ag films on Si(100) is the best candidate for this scenario.

*In summary*, we have studied the electron structure and morphology of Ag overlayers on Si(111) and Si(100) by means of XPS, UPS, LEED, STM and STS. Although, the XPS, UPS, LEED results show that low temperature Ag deposition favor a layer-by-layer like growth mode, a new unusual (√3x√3)R30° structure for 4-6 ML thick Ag films deposited onto a Si(111) (√3x√3)R30° surface at 130K has been observed. However, the STM study reveals a system of Ag 3D islands and √3Ag/Si 2D terraces in between. The low temperature deposition of 10 ML Ag on Si(100)-(2x1) yields an epitaxial growth of twinned Ag(111) films with enhanced conductivity. The resistivity of these films exhibits variable range hopping conductivity at low temperatures.

## ACKNOWLEDGMENTS

We are grateful to M. Henzler, H. Pfnür, G. Margaritondo, M. Grioni, W.-D. Schneider, C. Rüssel for helpful discussions. The authors have enjoyed stimulating discussions with W. Richter and B. Schröter. One of us (V.A.G.) is grateful to W. Richter for financial support during his stay in Jena by the DFG (Grant No. 436RUS 113/112/R) and Innovationskolleg INK 6/A1. We acknowledge a support from DFG through the SFB 196, RFBR (Grant No. 96-02-17532), RAS Program: New Materials and Structures (Grant 4.13).